# Planar Hall effect induced spin rectification effect and its strong impact on spin pumping measurements

Kang He[1], Jun Cheng[1], Man Yang[1], Yihui Zhang[3], Longqian Yu[1], Qi Liu[1], Liang Sun[1,2], Bingfeng Miao[1,2*], Canming Hu[3], Haifeng Ding[1,2]

[1]*National Laboratory of Solid State Microstructures and Department of Physics, Nanjing University, Nanjing, P.R. China*

[2]*Collaborative Innovation Center of Advanced Microstructures, Nanjing University, Nanjing, P.R. China*

[3]*Department of Physics and Astronomy, University of Manitoba, Winnipeg, Manitoba R3T 2N2, Canada*

*Corresponding author. E-mail: bfmiao@nju.edu.cn

## Abstract

Spin pumping is a technique widely used to generate the pure spin current and characterize the spin-charge conversion in various systems. The reversing sign of the symmetric Lorentzian charge current with respect to opposite magnetic field is generally accepted as the key criterion to identify its pure spin current origin. However, we herein find that the rectified voltage due to the planar Hall effect can exhibit similar spurious signal, complicating and even misleading the analysis. The distribution of microwave magnetic field and induction current has strong influence on the magnetic field symmetry and lineshape of the obtained signal. We further demonstrate a geometry where the spin-charge conversion and the rectified voltage can be readily distinguished with a straightforward symmetry analysis.

The generation, manipulation and detection of the pure spin currents are the central topics of spintronics in the last several decades [1-3]. Among various approaches, spin pumping offers an easy and versatile method to generate the pure spin currents, and it is not hampered by the resistance mismatch obstacle [4,5]. Upon the excitation of a microwave magnetic field with a suitable static magnetic field, the magnetization of a ferromagnet (FM) precesses meanwhile pumps a pure spin current into its adjacent layer [6-10]. The amplitude reaches its maximum at the ferromagnetic resonance (FMR) condition. Wherein, the pure spin current can be converted into a charge current and detected electronically in case the spin orbit coupling (SOC) exists. Spin pumping in combination with spin-charge conversion has been widely used to study the bulk SOC in heavy metals [7,11-14], spin-momentum locking of Rashba interface/surfaces [15-18] and topological insulators [19,20] etc.

In the inverse spin Hall effect (ISHE) in heavy metals, the inverse Rashba Edelstein effect (IREE) at surface/interfaces and the topological insulators, the converted charge current $\vec{j}_c$ can all be described by $\vec{j}_c \propto \vec{j}_s \times \vec{\sigma}$. Wherein, $\vec{j}_s$ represents the pure spin current and $\vec{\sigma}$ is the spin polarization direction. In the spin pumping measurements, $\vec{\sigma}$ is aligned by the magnetization orientation of FM in the FM/non-magnet (NM) heterostructures [6,11]. Therefore, $\vec{j}_c$ changes sign when the magnetization reverses. Specifically, the spin-charge conversion generated by spin pumping results in a voltage signal with symmetric Lorentzian lineshape at the FMR condition. And the voltage changes sign with the reversal of the FM magnetization [7,11-13,15-20]. It has also been well recognized that spin pumping is entangled with spurious contributions when the FM is conducting. For instance, its resistance oscillates with the precession of magnetization due to the anisotropic magnetoresistance (AMR) of ferromagnetic metals. And the coupling between dynamic resistance and the induction current along the stripe with the same frequency can result in a dc rectified voltage. This is the so called AMR induced spin rectification effect (SRE) [21]. Depending on the phase difference between the rf magnetic field and the induction current, SRE contains both symmetric and anti-symmetric Lorentzian contributions [21,22]. When the static magnetic field is rotated within the sample plane, spin pumping (AMR induced SRE) voltages are proportional to $\sin\phi_0$ ($\sin2\phi_0$) [21,22], respectively. Wherein, $\phi_0$ represents the angle between static magnetic field and voltage leads across the FM/NM heterostructure [Fig. 1(a)]. Therefore, AMR induced SRE disappears for $\phi_0=\pm90°$ in typical spin pumping induced spin-charge conversion measurements. At this specific geometry, a symmetric Lorentzian line-shaped voltage signal at FMR condition with $V(H)=-V(-H)$ is considered as the signature of spin pumping induced spin-charge conversion. This criterion has been established and widely used for spin pumping measurements in literature until now, including spin-charge conversion in Dirac semimetal [23], two-dimensional Rashba electron gas at interfaces [16,17,24], topological insulators [25,26], single layer graphene [27], superconductor [28], self-pumping of single Py [29-32] etc.

In this work, we demonstrate that this widely adopted standard for identifying the pure spin current origin of the measured signal is insufficient, as the importance of the planar Hall effect (PHE) is not appreciated. Depending on the geometry and the microwave frequency, PHE induced SRE can have similar behavior as the spin-charge conversion. Further, by placing a Py stripe in the gap between a signal line and the ground line of a coplanar waveguide (CPW), where the perpendicular magnetic rf field is dominant, we observe a voltage signal which is symmetric with magnetic field, i.e., $V(H)=V(-H)$ when $\phi_0=\pm90°$. The behavior is not compatible with either the spin pumping induced spin-charge conversion or the AMR induced SRE. Instead, it can be well explained by the PHE induced SRE. With increasing the Pt thickness of Py/Pt bilayers, the contribution of ISHE in Pt gradually dominates over the PHE induced SRE in Py. We further develop a quantitative method to separate the contributions of ISHE and SRE based on the symmetry analysis.

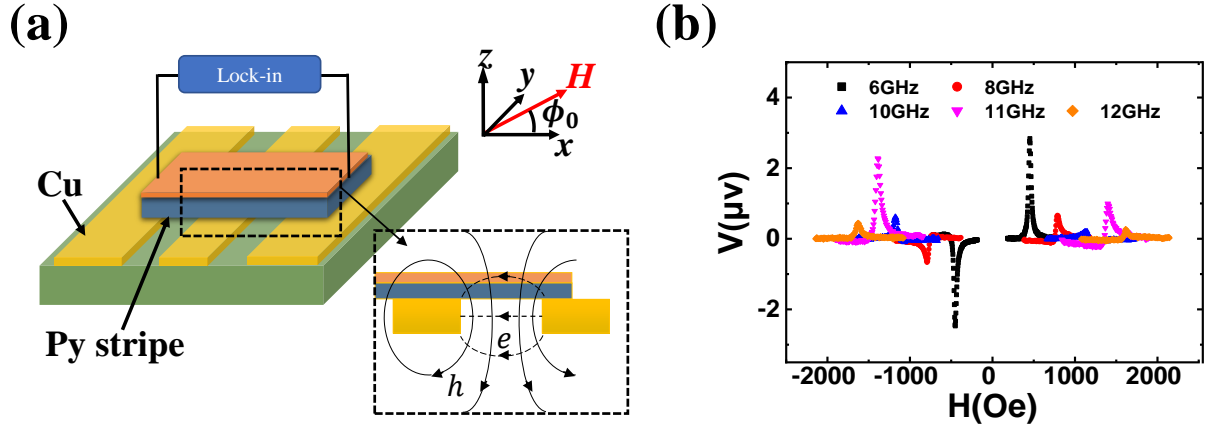

**FIG. 1**. (a) Schematic of the measurement geometry with a 2.5 mm×0.5 mm Py single stripe placed onto a commercial CPW. $\phi_0$ defines the angle between static magnetic field $H$ and the stripe. (b) $H$-dependent voltage signal of 10-nm thick Py stripe with microwave frequency varying from 6 GHz to 12 GHz. The magnetic field $H$ is applied transverse to the stripe along $\phi_0=\pm 90°$.

Py single layer and Py/Pt bilayers are deposited on the thermally oxidized Si substrates with dc magnetron sputtering, covered with 5-nm $SiO_2$ with rf magnetron sputtering. The samples are further patterned into stripes with lateral dimensions of 2 mm ×10 μm or 2.5 mm×0.5 mm using photolithography and lift-off techniques. And the films growth rates are calibrated with X-ray reflection. Figure 1 presents the schematic of the measurement geometry with a 2.5 mm×0.5 mm Py stripe placed onto a commercial CPW facing up. Microwave with the power of ~320 mW and different frequencies is fed into the CPW to excite the FMR of Py. To improve the signal-to-noise ratio, the microwave is further modulated by a 13.37-kHz TTL signal and the voltages along the stripes are measured with a lock-in amplifier. Rotatable magnetic field $H$ is applied within the sample plane with $\phi_0$ defined as the angle between $H$ and the stripe. And all measurements are performed at room temperature.

Figure 1(b) presents the $H$ dependent voltage signal of 10-nm thick Py stripe with $\phi_0=\pm 90°$. Voltage with dominant symmetric Lorentzian lineshape and anti-symmetric with the magnetic field, i.e., $V(H)=-V(-H)$, are observed when 6-GHz (black curve) and 8-GHz microwave (red curve) are applied. This feature is seemingly consistent with the spin pumping induced ISHE of the Py single layer [29-32], since AMR induced SRE disappears at this configuration. However, both the field symmetry and the lineshape of voltage change dramatically when the microwave frequency is increased to 11 GHz (magenta curve) and 12 GHz (brown curve). Voltage signal with the same sign albeit different magnitude emerges at the Py FMR condition for $H$ along $\phi_0=\pm 90°$. It is important to note that the spin pumping induced ISHE signal has only symmetric Lorentzian lineshape, and must change sign with reversing Py magnetization irrespective of the microwave frequency. Therefore, other contribution in the spin pumping measurement with metallic FM should be carefully explored before claiming its pure spin current origin.

Both the magnetic and electric field distribution above the CPW are three-dimensional and rather complex [Fig. 1(a) inset presents the schematic of magnetic field and electric field distribution above the CPW for one cross section] [33]. As many parameters in the expression for magnetic field and electric field are frequency dependent, their distributions are thus also frequency dependent. In addition, phase difference between the rf magnetic field and the induction current at different frequencies also has strong impact on the lineshape [22]. On the other hand, both the magnetic field (out-of-plane) and electric field (in-plane transverse) are one-dimensional in the gap between the signal line and ground line. For better understanding the physical origin of unexpected signal observed in Py single layer, we hereafter first focus on the spin pumping measurements with the stripe located in the gap between a signal line and the ground line of the CPW. We will qualitatively explain the observed feature presented in Fig. 1(b) later on.

Figure 2(a) presents the schematic of the measurement geometry with a Py(10 nm) stripe with the lateral dimension of 2 mm×10 μm. In this setup, Py stripe is exposed to an almost uniform microwave magnetic field along the *z*-direction, $h_z^{\rm rf}$. Surprisingly, unlike the anti-symmetric signal (8 GHz) with *H* presented in Fig. 1(b), the voltage of Py is symmetric with *H* and with almost identical amplitude at FMR condition for $\phi_0=\pm 90°$ [Fig. 2(b) black curve]. Under the in-plane transverse static magnetic field, $V(H)=V(-H)$ cannot be explained by either the spin pumping induced ISHE or the AMR induced SRE. Therefore, other physical mechanism should be explored. Traditionally, only the induction current along the stripe was considered in the AMR induced SRE. However, coexisting with the out-of-plane $h_z^{\rm rf}$, there is also an in-plane electric field of the same frequency (pointing from the signal line to the ground line of CPW) acting on the Py stripe [Fig. 2(a) inset]. The electric field of the CPW induces a dynamic current transverse to the stripe $j_y$ due to the Ohm's law which was seldom studied previously.

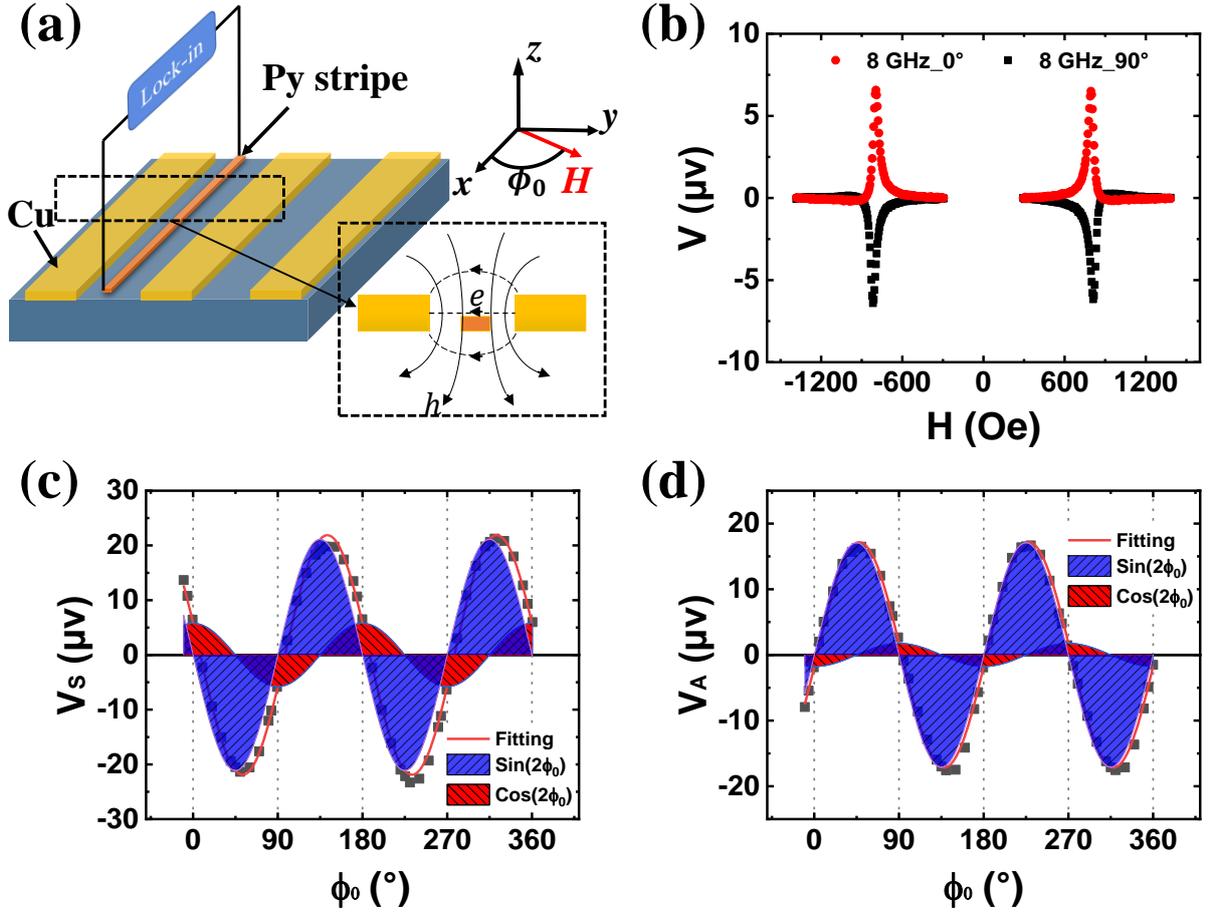

FIG. 2. (a) Schematic of the measurement geometry with a 2 mm×10 μm Py(10 nm) single stripe placed in the gap between a signal line and the ground line of the CPW. (b) $H$ dependent voltage signal of Py with $H$ along $\phi_0=\pm 90°$ (black curve), and $\phi_0=0°, 180°$ (red curve). The angular dependent (c) symmetric Lorentzian component $V_{sy}$ and (d) anti-symmetric Lorentzian component $V_{asy}$ for Py stripe. The black symbols are experimental data and the curve is the fitting. The blue/red shadowed area denotes the component of AMR/PHE induced rectification signal, respectively.

With the in-plane magnetization, the transverse ($y$-direction) current would result in a voltage along the stripe ($x$-direction) due to the PHE. PHE shares the same physical origin with AMR and is considered as the transverse version of the AMR. It has the in-plane angular dependence of $\sin 2\phi_0$. The rectified voltage due to PHE is proportional to d(PHE)/d$\phi_0$, i.e., $V_{SR}^{PHE} \propto \cos 2\phi_0$ [34-36]. This is consistent with the observed voltage signal with $V(H)=V(-H)$ for $\phi_0$=90° presented in Fig. 2(b). In addition, the two-fold symmetry of cos2$\phi_0$ predicts a sign change of $V_{SR}^{PHE}$ between $\phi_0$=0° and $\phi_0$=90°, while maintaining $V(H)=V(-H)$. Exactly the same feature is observed for the field dependent voltage at $\phi_0$=0° [red curve in Fig. 2(b)], evidencing the validity of our model. To further prove the importance of PHE induced SRE, we also perform the field dependent measurements for various $\phi_0$. At each certain direction, we decompose the voltage signal into symmetric Lorentzian component $V_S$ and antisymmetric Lorentzian component $V_A$:

$$V = V_S \frac{\Delta H^2}{(H-H_0)^2 + \Delta H^2} + V_A \frac{\Delta H(H-H_0)}{(H-H_0)^2 + \Delta H^2} \tag{1}$$

The angular dependent $V_S$ and $V_A$ for 10-nm Py stripe sitting in the gap of a CPW are presented in Figure 2(c) and 2(d) (black symbols), respectively. The data can be well fitted (red curve) by considering both AMR and PHE induced SRE:

$$V_{S(A)} = V_{SR}^{AMR} \sin 2\phi_0 + V_{SR}^{PHE} \cos 2\phi_0 \tag{2}$$

Where, the blue/red shadowed area denotes the component of AMR/PHE induced rectification signal, respectively. $V_{SR}^{AMR}$ and $V_{SR}^{PHE}$ are both two-fold symmetric and with a 45°-phase shift. In the gap of a CPW, we only consider the out-of-plane microwave magnetic field $h_z^{rf}$ and neglect the small in-plane component $h_x^{rf}$ and $h_y^{rf}$ [33,37]. The contributions of $h_x$ and $h_y$ for both the AMR and PHE induced rectifications are negligibly small, almost at the margin of error bar. The rather good fitting of experimental data with Eq. (2) also indicates negligibly small 'self-pumping' induced ISHE signal in the Py single layer as compared with the SRE. We note that the difference of the ratios between $V_{SR}^{AMR}$ and $V_{SR}^{PHE}$ for the symmetric [Fig. 2(c)] and the anti-symmetric component [Fig. 2(d)] is due to the different phase shift of $j_x$ and $j_y$ with respect to $h_z^{rf}$. However, the well-defined angular dependence of $\sin 2\phi_0$ and $\cos 2\phi_0$ consolidate that they come from the AMR and PHE, respectively. Equally important, both the PHE and AMR are proportional to the stripe length along the *x*-direction in our geometry. Thus, they could be in the similar order of magnitude as will be further discussed below.

With the understanding of the PHE induced SRE, we now provide a qualitative explanation of the ISHE-like signature $V(H) = -V(-H)$ of Py stripe placed onto the CPW at $\phi_0 = \pm 90°$ [6 GHz and 8GHz in Fig. 1(b)], where the AMR induced SRE disappears. When only $h_z^{rf}$ is present, the dynamic magnetic field is always perpendicular to the Py magnetization when it rotates within the sample plane. Thus, $V_{SR}^{PHE} \propto \cos 2\phi_0$ as presented in Fig. 2. When the excitation field is along the stripe (*x*-direction), only the component of $h_x^{rf}$ transverse to the Py magnetization contributes to the magnetization precession. Therefore, $V_{SR}^{PHE} \propto \cos 2\phi_0 \sin \phi_0$ for in-plane static *H* with in-plane $h_x^{rf}$. In this geometry, field antisymmetric voltage signal $V(H) = -V(-H)$ is expected from the solely PHE induced SRE, not necessarily the spin-charge conversion. When the Py stripe is placed onto the CPW [Fig. 1(a)], both $h_x^{rf}$ and $h_z^{rf}$ exert on the sample. The evolution of field symmetry and the lineshape of voltage curve with microwave frequency indicates the frequency dependent microwave magnetic field distribution and phase shift of magnetization precession [22,33]. When the $h_x^{rf}$ ($h_z^{rf}$) dominates, the signal follows $V(H) = -V(-H)$ [$V(H) = V(-H)$], respectively. The signal evolves when the relative contribution of $h_x^{rf}$ and $h_z^{rf}$ changes with frequency as observed in Fig. 1. This qualitatively explains the ISHE-like feature of Py single layer stripe. Our findings highlight the importance of a well-defined distribution of the microwave magnetic field. Specifically, $h_z^{rf}$ is most suitable for distinguish ISHE from SRE [geometry presented in Fig. 2(a)], as $V_{SR}^{AMR}$ disappears at $\phi_0 = \pm 90°$, and $V_{SR}^{PHE}$ [$V(H) = V(-H)$] and ISHE [$V(H) = -V(-H)$] reach their maximum magnitude but have different symmetries versus the magnetic field. Therefore, if opposite voltages between $\phi_0=\pm 90°$ is observed under $h_z^{rf}$ only, one could draw the conclusion that it indeed originates from the spin-charge conversion.

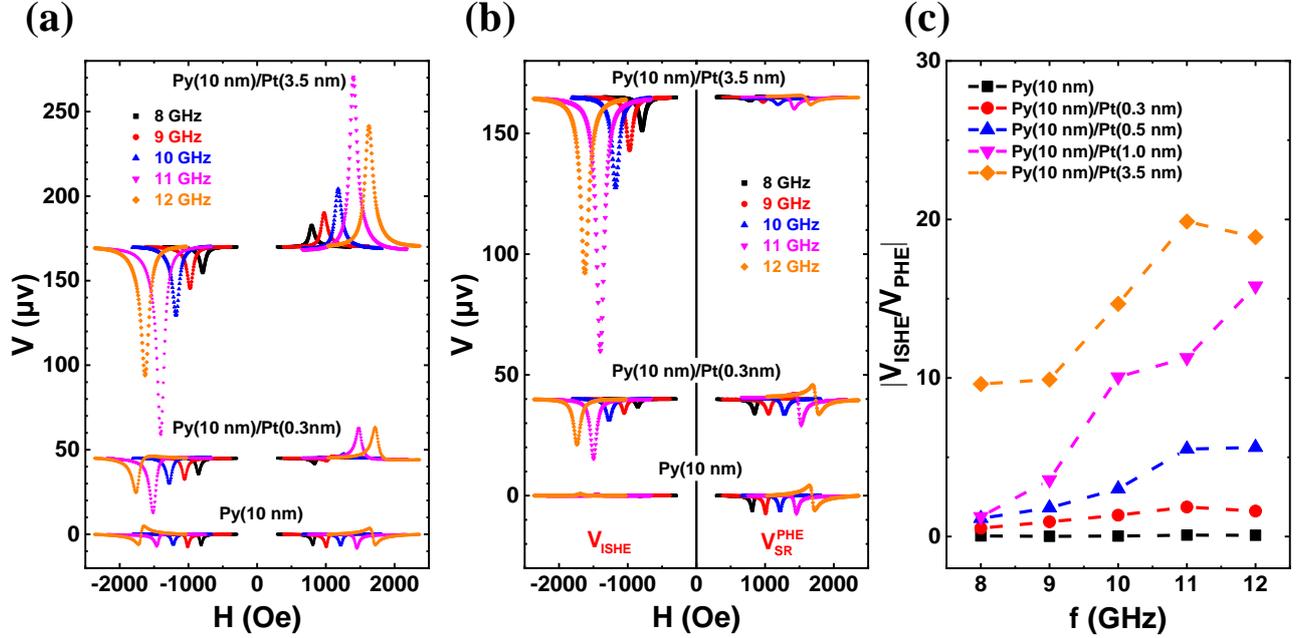

**FIG. 3**. (a) $H$ dependent voltage signal of Py(10 nm), Py(10 nm)/Pt(0.3 nm) and Py(10 nm)/Pt(3.5 nm) with microwave frequency varying from 8 GHz to 12 GHz. The magnetic field is applied along $\phi_0=\pm 90°$. (b) Extracted spin pumping induced ISHE contribution $V_{ISHE}$ (left) and PHE induced spin rectification contribution $V_{SR}^{PHE}$ (right) for Py(10 nm), Py(10 nm)/Pt(0.3 nm) and Py(10 nm)/Pt(3.5 nm). (c) Frequency dependent normalized ratio of $|V_{ISHE}/V_{PHE}|$ for Py(10 nm) single layer and Py(10 nm)/Pt($t_{Pt}$) bilayers.

To estimate the influence of the PHE induced SRE on the spin-charge conversion measurements with spin pumping, we perform the measurements for Py(10 nm)/Pt($t_{Pt}$) bilayers with various Pt thickness utilizing the $h_z^{rf}$ excitation [the same geometry as Fig. 2(a)]. Similar resonance fields of Py and Py/Pt($t_{Pt}$) indicate comparable magnetic property in all these samples. And we only compare the voltage signals for $\phi_0=\pm 90°$, where the AMR induced SRE vanishes. For Py/Pt bilayers, the ISHE signal $V_{\text{ISHE}}$ of Pt layer and the PHE rectification signal $V_{\text{SR}}^{\text{PHE}}$ of Py coexist. It is important to point out that there is also spin Hall magnetoresistance (SMR) in the Py/Pt bilayer, due to the combination of the spin Hall effect and ISHE of Pt and magnetic dependent scattering at Py-Pt interface [38,39]. Although SMR and AMR have different physical origins with different symmetries, their in-plane angular dependences of FM magnetization are the same [38,39]. Therefore, the $V_{\text{SR}}^{\text{PHE}}$ due to SMR or AMR is additive and in-distinguishable with in-plane FM magnetization. In this work, we would not separate the SMR and AMR contributions in $V_{\text{SR}}^{\text{PHE}}$. As it has been presented in Fig. 2(b), $V_{\text{SR}}^{\text{PHE}}$ is dominant for the Py single layer ($t_{\text{Pt}}=0$ nm). The signals maintain the symmetry of $V(H)=V(-H)$, although the detailed lineshape changes largely between 8 GHz and 12 GHz [Fig. 3(a)]. In Py/Pt(3.5 nm), the $V_{\text{ISHE}}$ of Pt is dominant and the entire signal changes sign with reversing the magnetic field direction for microwave between 8 GHz and 12 GHz. The slight difference between the amplitudes at positive and negative fields is due to the minor contribution from $V_{\text{SR}}^{\text{PHE}}$ of the Py layer. With increasing the Pt thickness, the signal gradually evolves from $V(H)=V(-H)$ to $V(H)=-V(-H)$ as the contribution of $V_{\text{ISHE}}$ enhances and becomes dominant at large Pt thickness. Interestingly, for Py/Pt(0.3 nm) at 9 GHz, one observe a sizable signal at negative field dominated by symmetric Lorentzian lineshape and a much weaker voltage at positive field dominated by anti-symmetric Lorentzian lineshape. Because $V_{\text{ISHE}}$ and $V_{\text{SR}}^{\text{PHE}}$ are additive at the negative field, they are subtractive at the positive field. And $V_{\text{ISHE}}$ has only symmetric Lorentzian component, while $V_{\text{SR}}^{\text{PHE}}$ has both symmetric and anti-symmetric Lorentzian component depending on the phase shift between the magnetization precession and induction current along the y-direction. When $V_{\text{ISHE}}$ and the symmetric Lorentzian component of $V_{\text{SR}}^{\text{PHE}}$ have similar magnitude, the residual signal at positive field is mainly the anti-symmetric Lorentzian component of $V_{SR}^{PHE}$.

From the opposite field dependence of $V_{ISHE}$ and $V_{SR}^{PHE}$, we can readily extract both contributions at $\phi_0=\pm90°$ via:

$$V_{ISHE} = \frac{V_S(H)-V_S(-H)}{2}, V_{SR}^{PHE} = \frac{V(H)+V(-H)}{2} \quad (3)$$

The results are presented at Fig. 3(b) with extracted ISHE contribution (left) and PHE rectification contribution (right) for Py(10 nm), Py(10 nm)/Pt(0.3 nm) and Py(10 nm)/Pt(3.5 nm). $V_{SR}^{PHE}$ decreases with increasing Pt thickness due to the shunting effect. And $V_{ISHE}$ increases with Pt thickness in the thin range owing to the ISHE and the spin diffusion effect in Pt [13,40,41]. We also replot the frequency dependent normalized $\left|\frac{V_{ISHE}}{V_{SR}^{PHE}}\right|$ for Py/Pt($t_{Pt}$) bilayers in Fig. 3(c). Here, $\left|V_{SR}^{PHE}\right|$ is the amplitude of PHE induced SRE which includes both the symmetric and anti-symmetric Lorentzian components. $\left|\frac{V_{ISHE}}{V_{SR}^{PHE}}\right|$ increases with $t_{Pt}$ and frequency, suggesting that high frequency excitation is more reliable for exploring spin-charge conversion with spin pumping. We note that although the SRE can be suppressed if ferromagnetic insulator yttrium iron garnet (YIG) is chosen as the spin current source, the necessary heating process is detrimental for many soft two-dimensional topological materials [42-44]. Further, high quality YIG largely relies on the gadolinium gallium garnet substrate [42-44]. On the other hand, metallic FMs can be deposited on different substrates/underlayers without special treatment. Thus, they have been extensively used as spin current injectors in various studies of spin-charge conversions. The complete understanding of SRE in metallic FM is therefore pivotal to investigate spin current with spin pumping technique.

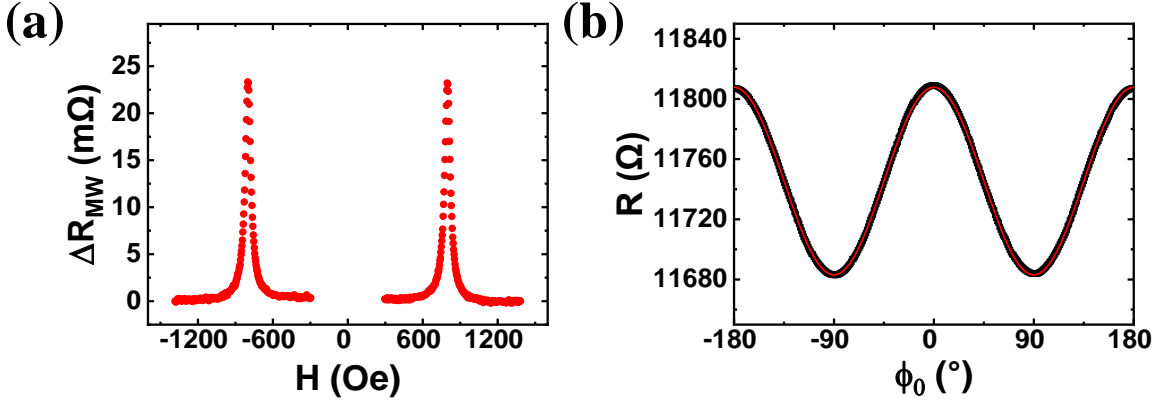

**FIG. 4**. (a) Microwave photoresistance $\Delta R_{MW}$ for Py(10 nm) with 8-GHz microwave. (b) Angular dependent resistance of Py (10 nm) with fitting (red curve).

Lastly, we provide the estimations of the in-plane longitudinal and transverse induction current density, in-plane electric field, and out-of-plane magnetic field when the sample sits in-between the signal line and ground line of a CPW [geometry in Fig. 2(a)]. The discussions are perform for Py(10 nm) under 8-GHz microwave and applicable for every sample. At the FMR, the magnetization precession alters the angle of the magnetization with respect to the dc current, resulting in a change of the time-averaged AMR. This is termed as the microwave photoresistance $\Delta R_{MW}$. For $\phi_0=90°$, it can be given by:

$$\Delta R_{MW} = \frac{R_A}{2} \alpha_1^2 \frac{\Delta H^2}{(H-H_0)^2 + \Delta H^2} \tag{4}$$

Here, $R_A$ is the AMR value, and $\alpha_1$ is the amplitude of in-plane precession angle of the FM magnetization [45,46]. Fig. 4(a) presents the magnetic field-dependent $\Delta R_{MW}$ for Py(10 nm) at 8 GHz. With $R_A = 124.68\ \Omega$ obtained in the angular dependent resistance [Fig. 4(b)], we obtain $\alpha_1 = 1.103°$. We can further estimate the out-of-plane rf magnetic field $h_z^{rf}$ through $\alpha_1 = \dfrac{h_z^{rf}}{\alpha_G (2H_0 + M_{eff})}$ [45,46]. With the Gilbert damping factor $\alpha_G = 0.011$ and effective magnetization $M_{eff} = 9189\,\text{Oe}$ obtained from frequency dependent half line-width and resonance field of Py [Fig. 3], we get $h_z^{rf} = 2.3\,\text{Oe}$. The amplitude of AMR induced SRE can be described by $\left|V_{SR}^{AMR}\right| = \frac{1}{2}\left|R_A I_x\right|$, where $I_x$ is longitudinal induction current [45,46]. Accordingly, the amplitude of PHE induced SRE can be described by $\left|V_{SR}^{PHE}\right| = \frac{1}{2}\left|R_A I_y\right|$, because both PHE and AMR use the same voltage lead with orthogonal current. We therefore obtain the longitudinal induction current density $j_x = 226.2\,\mu\text{A}/\mu\text{m}^2$, and transverse induction current density $j_y = 50.4\,\mu\text{A}/\mu\text{m}^2$. And the transverse electric field $E_y = 29.4\,\mu\text{V}/\mu\text{m}$ is obtained through Ohm's law. Our estimations of the $j_x$ and $h_z^{rf}$ are reasonably consistent with the limited reports in literature [21,47], although the detailed dimensions of CPW and sample even lead configuration and wiring conditions of a particular device may have influence [22].

In summary, we have studied the influence of rectified voltage from PHE and microwave electromagnetic field distribution in the spin pumping measurements of Py single layer and Py/Pt bilayers. When the in-plane longitudinal microwave magnetic field is present, SRE from PHE and spin-charge conversion have the same magnetic field symmetry. The spurious signal from PHE may lead to incorrect conclusion of pure spin current origin. In addition, we also demonstrate a geometry with well-defined out-of-plane microwave magnetic field, where the PHE induced SRE and spin-pumping induced ISHE can be readily distinguished. Our findings also suggest a revisit of a few important controversies of spin-charge conversion may be necessary, where the magnetic metals were used as spin pumping sources.

**Acknowledgement**

This work was supported by the National Key R&D Program of China (Grant No. 2018YFA0306004 and 2017YFA0303202), the National Natural Science Foundation of China (Grants No. 51971110, No. 11974165, No. 11734006, No. 11727808), Natural Science Foundation of Jiangsu

Province (Grant No. BK20190057). C.-M. H. is funded by the NSERC Discovery Grants and the NSERC Discovery Accelerator Supplements.**References:**
[1] S. D. Bader and S. S. P. Parkin, Annu. Rev. Cond. Matter Phys. **1**, 71 (2010).
[2] J. Sinova, S. O. Valenzuela, J. Wunderlich, C. H. Back, and T. Jungwirth, Rev. Mod. Phys. **87**, 1213 (2015).
[3] Y. Niimi and Y. Otani, Rep. Prog. Phys. **78**, 124501 (2015).
[4] S. Watanabe, K. Ando, K. Kang, S. Mooser, Y. Vaynzof, H. Kurebayashi, E. Saitoh, and H. Sirringhaus, Nat. Phys. **10**, 308 (2014).
[5] C. H. Du, H. L. Wang, Y. Pu, T. L. Meyer, P. M. Woodward, F. Y. Yang, and P. C. Hammel, Phys. Rev. Lett. **111**, 247202 (2013).
[6] Y. Tserkovnyak, A. Brataas, and G. E. W. Bauer, Phys. Rev. Lett. **88**, 117601 (2002).
[7] E. Saitoh, M. Ueda, H. Miyajima, and G. Tatara, Appl. Phys. Lett. **88**, 182509 (2006).
[8] Y. Otani, M. Shiraishi, A. Oiwa, E. Saitoh, and S. Murakami, Nat. Phys. **13**, 829 (2017).
[9] S. M. Watts, J. Grollier, C. H. van der Wal, and B. J. van Wees, Phys. Rev. Lett. **96**, 077201 (2006).
[10] B. Heinrich, C. Burrowes, E. Montoya, B. Kardasz, E. Girt, Y.-Y. Song, Y. Sun, and M. Wu, Phys. Rev. Lett. **107**, 066604 (2011).
[11] O. Mosendz, J. E. Pearson, F. Y. Fradin, G. E. W. Bauer, S. D. Bader, and A. Hoffmann, Phys. Rev. Lett. **104**, 046601 (2010).
[12] A. Azevedo, L. H. Vilela-Leão, R. L. Rodríguez-Suárez, A. F. Lacerda Santos, and S. M. Rezende, Phys. Rev. B **83**, 144402 (2011).
[13] X. Tao, Q. Liu, B. Miao, R. Yu, Z. Feng, L. Sun, B. You, J. Du, K. Chen, S. Zhang *et al.*, Sci. Adv. **4**, eaat1670 (2018).
[14] V. Castel, N. Vlietstra, B. J. van Wees, and J. Ben Youssef, Phys. Rev. B **90**, 214434 (2014).
[15] J. C. R. Sánchez, L. Vila, G. Desfonds, S. Gambarelli, J. P. Attané, J. M. De Teresa, C. Magén, and A. Fert, Nat. Commun. **4**, 2944 (2013).
[16] E. Lesne, Y. Fu, S. Oyarzun, J. C. Rojas-Sánchez, D. C. Vaz, H. Naganuma, G. Sicoli, J. P. Attané, M. Jamet, E. Jacquet *et al.*, Nat. Mater. **15**, 1261 (2016).
[17] Q. Song, H. Zhang, T. Su, W. Yuan, Y. Chen, W. Xing, J. Shi, J. Sun, and W. Han, Sci. Adv. **3**, e1602312 (2017).
[18] R. Yu, B. Miao, Q. Liu, K. He, W. Xue, L. Sun, M. Wu, Y. Wu, Z. Yuan, and H. Ding, Phys. Rev. B **102**, 144415 (2020).
[19] Y. Shiomi, K. Nomura, Y. Kajiwara, K. Eto, M. Novak, K. Segawa, Y. Ando, and E. Saitoh, Phys. Rev. Lett. **113**, 196601 (2014).
[20] M. Jamali, J. S. Lee, J. S. Jeong, F. Mahfouzi, Y. Lv, Z. Zhao, B. K. Nikolić, K. A. Mkhoyan, N. Samarth, and J.-P. Wang, Nano Lett. **15**, 7126 (2015).
[21] Y. S. Gui, N. Mecking, X. Zhou, G. Williams, and C. M. Hu, Phys. Rev. Lett. **98**, 107602 (2007).
[22] M. Harder, Z. X. Cao, Y. S. Gui, X. L. Fan, and C. M. Hu, Phys. Rev. B **84**, 054423 (2011).
[23] W. Yanez, Y. Ou, R. Xiao, J. Koo, J. T. Held, S. Ghosh, J. Rable, T. Pillsbury, E. G. Delgado, K. Yang *et al.*, Phys. Rev. Applied **16**, 054031 (2021).
[24] P. Noël, F. Trier, L. M. Vicente Arche, J. Bréhin, D. C. Vaz, V. Garcia, S. Fusil, A. Barthélémy, L. Vila, M. Bibes *et al.*, Nature **580**, 483 (2020).
[25] J. B. S. Mendes, M. Gamino, R. O. Cunha, J. E. Abrão, S. M. Rezende, and A. Azevedo, Phys. Rev. Mater. **5**, 024206 (2021).
[26] J. B. S. Mendes, O. Alves Santos, J. Holanda, R. P. Loreto, C. I. L. de Araujo, C.-Z. Chang, J. S. Moodera, A. Azevedo,
15

and S. M. Rezende, Phys. Rev. B **96**, 180415(R) (2017).

[27] J. B. S. Mendes, O. Alves Santos, T. Chagas, R. Magalhães-Paniago, T. J. A. Mori, J. Holanda, L. M. Meireles, R. G. Lacerda, A. Azevedo, and S. M. Rezende, Phys. Rev. B **99**, 214446 (2019).

[28] K.-R. Jeon, C. Ciccarelli, H. Kurebayashi, J. Wunderlich, L. F. Cohen, S. Komori, J. W. A. Robinson, and M. G. Blamire, Phys. Rev. Applied **10**, 014029 (2018).

[29] A. Tsukahara, Y. Ando, Y. Kitamura, H. Emoto, E. Shikoh, M. P. Delmo, T. Shinjo, and M. Shiraishi, Phys. Rev. B **89**, 235317 (2014).

[30] A. Azevedo, R. O. Cunha, F. Estrada, O. Alves Santos, J. B. S. Mendes, L. H. Vilela-Leão, R. L. Rodríguez-Suárez, and S. M. Rezende, Phys. Rev. B **92**, 024402 (2015).

[31] O. Gladii, L. Frangou, A. Hallal, R. L. Seeger, P. Noël, G. Forestier, S. Auffret, M. Rubio-Roy, P. Warin, L. Vila *et al.*, Phys. Rev. B **100**, 174409 (2019).

[32] A. Azevedo, O. A. Santos, R. O. Cunha, R. Rodríguez-Suárez, and S. M. Rezende, Appl. Phys. Lett. **104**, 152408 (2014).

[33] R. N. Simons and R. K. Arora, IEEE Trans. Microwave Theory Tech. **30**, 1094 (1982).

[34] J. C. Rojas-Sánchez, M. Cubukcu, A. Jain, C. Vergnaud, C. Portemont, C. Ducruet, A. Barski, A. Marty, L. Vila, J. P. Attané *et al.*, Phys. Rev. B **88**, 064403 (2013).

[35] L. Chen, F. Matsukura, and H. Ohno, Nat. Commun. **4**, 2055 (2013).

[36] M. Harder, Y. Gui, and C.-M. Hu, Phys. Rep. **661**, 1 (2016).

[37] L. Bai, Z. Feng, P. Hyde, H. F. Ding, and C.-M. Hu, Appl. Phys. Lett. **102**, 242402 (2013).

[38] H. Nakayama, M. Althammer, Y. T. Chen, K. Uchida, Y. Kajiwara, D. Kikuchi, T. Ohtani, S. Geprägs, M. Opel, S. Takahashi *et al.*, Phys. Rev. Lett. **110**, 206601 (2013).

[39] N. Vlietstra, J. Shan, V. Castel, B. J. van Wees, and J. Ben Youssef, Phys. Rev. B **87**, 184421 (2013).

[40] B. B. Singh, K. Roy, J. A. Chelvane, and S. Bedanta, Phys. Rev. B **102**, 174444 (2020).

[41] H. Wang, Y. Xiao, M. Guo, E. Lee-Wong, G. Q. Yan, R. Cheng, and C. R. Du, Phys. Rev. Lett. **127**, 117202 (2021).

[42] H. L. Wang, C. H. Du, Y. Pu, R. Adur, P. C. Hammel, and F. Y. Yang, Phys. Rev. B **88**, 100406(R) (2013).

[43] Y. Sun, Y.-Y. Song, H. Chang, M. Kabatek, M. Jantz, W. Schneider, M. Wu, H. Schultheiss, and A. Hoffmann, Appl. Phys. Lett. **101**, 152405 (2012).

[44] H. Chang, P. Li, W. Zhang, T. Liu, A. Hoffmann, L. Deng, and M. Wu, IEEE Magn. Lett. **5**, 6700104 (2014).

[45] N. Mecking, Y. S. Gui, and C. M. Hu, Phys. Rev. B **76**, 224430 (2007).

[46] Z. Feng, J. Hu, L. Sun, B. You, D. Wu, J. Du, W. Zhang, A. Hu, Y. Yang, D. M. Tang *et al.*, Phys. Rev. B **85**, 214423 (2012).

[47] M. Obstbaum, M. Härtinger, H. G. Bauer, T. Meier, F. Swientek, C. H. Back, and G. Woltersdorf, Phys. Rev. B **89**, 060407(R) (2014).